\documentclass[epj,nopacs]{svjour}
\usepackage{graphicx}
\begin{document}
\title{Hadronic three-body decays of light vector mesons}
\author{S. Leupold\inst{1} \and M.F.M. Lutz\inst{2}
}                     
%
%
\institute{Institut f\"ur Theoretische Physik, Johann Wolfgang Goethe-Universit\"at Frankfurt, Germany
\and GSI, Planckstrasse 1,  D-64291 Darmstadt, Germany}
\date{\today}
%
\abstract{
The decays of light vector mesons into three pseudoscalar mesons are calculated to leading
order in the recently proposed counting scheme that is based on the hadrogenesis conjecture. 
Fully differential as well as integrated decay widths are presented.
Since the required parameters have been fixed by other processes, the considered
three-body decays are predictions of the presented approach. The decay width of the omega meson
into three pions agrees very well with experiment. The partial decay widths
of the $K^*$ into its three $K \pi \pi$ channels are predicted.
\PACS{
      {PACS-key}{discribing text of that key}   \and
      {PACS-key}{discribing text of that key}
     } 
} 
\maketitle
\section{Introduction}
\label{sec:intro}

One of the open challenges of QCD is to develop a systematic scheme for the calculation of hadronic reactions and decays.
In the sector of mesons made out of light quarks, chiral perturbation theory \cite{Gasser:1984gg,Scherer:2002tk}
provides such a systematic approach. However, it is restricted to energies where no other mesons than the Goldstone bosons 
participate as active degrees of freedom. It is therefore important to construct effective 
field theories with additional degrees of freedom. In particular, in the framework of the hadrogenesis conjecture 
\cite{Lutz:2003fm} such an extension is 
mandatory since the pseudoscalar and vector mesons together with the baryon octet and decuplet are regarded as the
preselected states which build up the other hadrons --- at least in the low-energy regime.

Recently we have suggested in \cite{Lutz:2008km} a counting scheme for flavor-SU(3) systems of Goldstone bosons and light 
vector mesons. The quest is to have a scheme at hand which allows in particular for decays of vector mesons into pseudoscalar states. The 
masses of pseudoscalar ($P$) and vector mesons ($V$) and all involved momenta are treated as soft, i.e.\ $m_P$, $m_V \sim Q$, 
where $Q$ is a typical momentum.  Such an assumption may be justified in QCD with a large number of colors $N_c$, which implies 
that the chiral symmetry breaking scale  is much larger than the vector meson masses, i.e. $4\pi f_\pi \gg m_V$. Consequently,  
all chirally covariant derivatives $D_\mu$ are $\sim Q$, no matter whether they act on pseudoscalar or vector
mesons. This is in contrast to the approach where vector mesons are treated as heavy matter fields \cite{Jenkins:1995vb}. The 
latter formalism does not allow for a systematic treatment of processes where the number of vector mesons changes, like 
e.g.\ in vector-meson decays. Our scheme is also different from the one presented in \cite{Ecker:1988te,Ecker:1989yg} where 
all derivatives are treated as soft, but the masses of the vector mesons are treated as heavy. Strictly speaking, such an approach is
restricted to energies below the vector meson masses. Nonetheless, resonance saturation models based on
large-$N_c$ considerations  have been built upon the works
\cite{Ecker:1988te,Ecker:1989yg} and have been applied also to processes at higher energies,
see e.g.\ \cite{RuizFemenia:2003hm} and references therein.

For the present purpose it is only important to recall from \cite{Lutz:2008km} that loops are suppressed
which is a generic consequence of large-$N_c$ considerations \cite{tHooft:1973jz}. It is important to note 
that the vector mesons are represented by tensor fields in our approach. The chosen representation cannot change 
the physical results, but contact terms (e.g.\ four-point interactions) can look different in different representations 
and different orders in a given counting scheme might be attributed to them.

It has been demonstrated in \cite{Lutz:2008km} that the proposed scheme can be used to describe hadronic
and electromagnetic two-body decays of vector mesons. In the present work we explore some further consequences of this 
framework by calculating to leading order the decays of vector mesons into three pseudoscalar mesons. It will turn out that 
all necessary parameters for the leading-order calculation have already been determined from the two-body decays. Therefore, 
our calculations can be tested against the available experimental data and we can provide predictions for processes which
were not measured yet or not measured in a fully differential way.

In the following we will study the process $\omega \to 3\pi$ and the three decay branches of
$K^* \to K \, \pi \, \pi$. There are no data for any differential widths of these 
three-body decays available. The partial decay width for the omega decay is measured \cite{Yao:2006px} 
whereas for the $K^*$ decays only upper limits exist \cite{Jongejans:1977ty}. The omega decay
has been studied by several groups, see e.g.\ \cite{Klingl:1996by} and references therein.
Typically, however, the corresponding coupling constants have been fitted to the partial decay width
of the process $\omega \to 3\pi$. In our approach we have determined the required coupling constants from
two-body processes and we have a counting scheme which tells us that we have covered all relevant vertices.
An approach which is on the technical level close to ours is the one presented in \cite{RuizFemenia:2003hm}.
Also there tensor fields have been used to represent the vector mesons. However, the counting scheme is
different since vector meson masses are treated as heavy in \cite{RuizFemenia:2003hm}. Therefore, also
the results cannot be compared easily. For the decays of the $K^*$ the literature is scarce.
In \cite{Boal:1976eg} various models including vector-meson dominance and also contact interactions
(see discussion below in sect. \ref{sec:kinemat}) have been applied to the three-body decays of the $K^*$.
We will comment on the work of \cite{Boal:1976eg} when we present our results below. In contrast to previous works 
differential decay widths for the considered processes will be presented. To the best of our knowledge this issue 
has not been studied at all by other theory groups.

The work is structured in the following way: In the next section we present the relevant part of our
leading-order Lagrangian and provide some general formulae important for three-body decays.
In sect.\ \ref{sec:om3pi} we calculate the decay of the omega meson into three pions. We will
present the fully differential decay width (Dalitz plot) as well as integrated quantities thereof.
The corresponding calculations for the decay channels of the $K^*$ into one kaon and two pions
are presented in sect.\ \ref{sec:dec-KsKpipi}. Finally in sect.\ \ref{sec:sum} we summarize our
results.

\section{Three-body decays}
\label{sec:kinemat}

\begin{figure}[ht]
\centering
\resizebox{0.3\textwidth}{!}{%
    \includegraphics{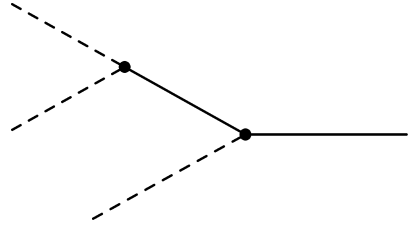}
}
  \caption{Feynman diagram which contributes in leading order to the decay of a vector meson (solid line)
into three pseudoscalar mesons (dashed lines).}
  \label{fig:vto3p}
\end{figure}

We compute the decays of light vector mesons into three pseudoscalar mesons. In general, such a calculation involves (a) 
tree diagrams with three-point vertices, (b) a tree diagram with a four-point vertex and (c) loops. We shall show now that 
only case (a) is relevant at leading order in the scheme proposed in  \cite{Lutz:2008km}: For the considered decays the 
leading-order calculation is of second order. Loops are suppressed  contributing at fourth order \cite{Lutz:2008km}. 
In principle, it is conceivable that a four-point vertex with one vector and three pseudoscalar states could enter at 
leading, i.e.\ second, order. However, within the tensor representation, which we use, one needs at least four derivatives to 
construct such a vertex. Thus it has at least order four according to the counting rules of \cite{Lutz:2008km}.  
Hence at the leading-order calculation, on which we concentrate in the following, one only needs three-point vertices. Since 
a three-point vertex with three pseudoscalar states does not exist, the decay process happens in two steps: The initial 
vector meson decays into another vector meson and one pseudoscalar state. The emerging vector meson finally decays into 
two pseudoscalar states. The generic diagram is depicted in fig.\ \ref{fig:vto3p}.

We recall from \cite{Lutz:2008km} the relevant part of the leading-order Lagrangian:
\begin{eqnarray}
{\mathcal L} &=& - \frac{h_{A}}{16 \, f}\,\epsilon^{\mu\nu\alpha\beta}\,
{\rm tr}\,\left\{
\left(V_{\mu \nu}\,(\partial^\tau V_{\tau \alpha})+(\partial^\tau V_{\tau \alpha})\,V_{\mu \nu} \right)
\,\partial_\beta \Phi\right\}
\nonumber\\
&& {} +i\,\frac{m_V\, h_{P}}{8 \,f^2}\,
{\rm tr}\,\left\{V^{\mu\nu}\,\partial_\mu \Phi \, \partial_\nu \Phi \right\}
\nonumber\\
&& {} - \frac{b_A}{16\, f} \, \epsilon^{\mu\nu\alpha\beta}\, {\rm tr}\,
\left\{\left[V_{\mu\nu},\,V_{\alpha \beta}\right]_+ \, [\Phi,\chi_0]_+ \right\} \,.
\label{interaction-tensor}
\end{eqnarray}
with the conventions
\begin{eqnarray}
&& V_{\mu \nu} = \left(\begin{array}{ccc}
\rho^0_{\mu \nu}+\omega_{\mu \nu} &\sqrt{2}\,\rho_{\mu \nu}^+&\sqrt{2}\,K_{\mu \nu}^+\\
\sqrt{2}\,\rho_{\mu \nu}^-&-\rho_{\mu \nu}^0+\omega_{\mu \nu}&\sqrt{2}\,K_{\mu \nu}^0\\
\sqrt{2}\,K_{\mu \nu}^- &\sqrt{2}\,\bar{K}_{\mu \nu}^0&\sqrt{2}\,\phi_{\mu \nu}
\end{array}\right) \,,
\label{def-fields-vector}
\\
&&  \Phi =\left(\begin{array}{ccc}
\pi^0+\frac{1}{\sqrt{3}}\,\eta &\sqrt{2}\,\pi^+&\sqrt{2}\,K^+\\
\sqrt{2}\,\pi^-&-\pi^0+\frac{1}{\sqrt{3}}\,\eta&\sqrt{2}\,K^0\\
\sqrt{2}\,K^- &\sqrt{2}\,\bar{K}^0&-\frac{2}{\sqrt{3}}\,\eta
\end{array}\right) \,,
\label{def-fields-scalar}
\\
&& \chi_0 = 
\left(
\begin{array}{ccc}
m_\pi^2 & 0 & 0\\
0 & m_\pi^2 & 0 \\
0 & 0 & 2 \,m_K^2 - m_\pi^2
\end{array}
\right)
\,.
\label{GMOR}
\end{eqnarray}

In (\ref{interaction-tensor}) the quantity $f$ denotes the pion-decay constant in the chiral limit.
The quantity $m_V$ has been introduced for convenience to render the coupling constants in  (\ref{interaction-tensor}) dimensionless.
The combination $m_V \, h_P/f^2$ has been fixed in \cite{Lutz:2008km} from the decays of vector mesons
into two pseudoscalar states. The leading-order Lagrangian of our approach has only one such term.
Hence it predicts a universal value for the coupling constant for all these decays, i.e.\ for
$\rho \to 2\pi$, $K^* \to \pi \, K$ and $\phi \to K \, \bar K$. Indeed, the values for $h_P$ obtained
from fits to the experimental decay widths agree with each other within $\pm$10\%.
The values for $h_A$ and $b_A$ have been determined from the radiative decays of vector mesons \cite{Lutz:2008km}. Here, 
it turned out that one universal vertex would be insufficient to describe all the radiative two-body decays of the vector 
meson nonet. Indeed, our leading-order Lagrangian yields two terms ---one flavor symmetric, one flavor breaking --- which 
are then sufficient to describe the data \cite{Lutz:2008km}. We regard this as a support of our approach. 

The only decays which are allowed by energy-momen\-tum conservation and where our Lagrangian yields
non-vanishing couplings at leading order are $\omega \to 3\pi$ and $K^* \to 2\pi + K$.
Note that in leading order the $\phi$ is purely strange and OZI suppression \cite{Okubo:1963fa,Zweig:1964,Iizuka:1966fk} is perfect.
Hence we do not provide results for the OZI suppressed decay branch $\phi \to 3\pi$.

In the following we use the values
\begin{eqnarray}
  && f = 90 \, \mbox{MeV,} \quad m_V = 776 \, \mbox{MeV,} \nonumber \\
  && h_P = 0.304 \,, \quad h_A = 2.1 \,, \quad b_A = 0.27 \,.
  \label{eq:paramval}
\end{eqnarray}
Note that the precise value for $h_P$ emerges from a fit to the electromagnetic pion
form factor \cite{Leupold:2008is}. For the two-body decay of the $K^*$ this value
for $h_P$ implies a partial width of $50.9$ MeV, in excellent agreement with data \cite{Yao:2006px}.
In \cite{Lutz:2008km} we have deduced for $h_P$ a value of $0.29 \pm 0.03$ which agrees with the one
given in (\ref{eq:paramval}) within the error bars. This result of \cite{Lutz:2008km} emerged from
a simultaneous fit to the decay widths $\rho \to \pi\, \pi$, $K^* \to K \, \pi$ and $\phi \to K \, \bar K$.
For the processes of interest in the present context the $\phi$ decay does not enter. Hence we use
a value for $h_P$ which agrees well with the processes $\rho \to \pi\, \pi$ and $K^* \to K \, \pi$ relevant 
for the decay chains depicted in fig.\ \ref{fig:vto3p}. 

We consider the decay of a state $V$ with momentum $p$ into three states with momenta
$p_1$, $p_2$ and $p_3$.  According to \cite{Yao:2006px} the double differential decay rate 
\begin{eqnarray}
  \label{eq:diffgam}
&&  \frac{d^2\Gamma_{V\to 1 2 3}}{dm_{12}^2 \, dm_{23}^2} =
  \frac{1}{(2\pi)^3}\, \frac{P}{32\, M^3}
  \, \Big| C_{V\to 1\,2\,3} \Big|^2  \,,
\nonumber\\
&& P  =   - \frac13 \, \epsilon_{\mu\nu\alpha\beta} \, p^\mu \, p_1^\nu \, p_2^\alpha \,
       {\epsilon_{\bar \mu \bar \nu \bar \alpha}}^\beta \, p^{\bar\mu} \, p_1^{\bar\nu}\, p_2^{\bar\alpha} \,, 
\end{eqnarray}
is determined by a matrix element $C_{V\to 1\,2\,3} $ and the phase-space factor $P$. 
In (\ref{eq:diffgam})  we introduced the following variables
\begin{eqnarray}
&& m_{ij}^2 = (p_i+p_j)^2 \,, \qquad \qquad M^2 =p^2  \,,
\nonumber\\
&& q_1^2  =  m_{23}^2 \,, \qquad \qquad \qquad \quad  \;q_3^3  =  m_{12}^2 \,,
\nonumber \\
&& q_2^2  = m_{13}^2= p^2 - m_{12}^2 - m_{23}^2 + p_1^2+p_2^2+p_3^2  \,. 
\label{eq:two-body-var}
\end{eqnarray}
The additional quantities $q_i^2$ are introduced for later convenience.

\section{Decay $\omega \to 3\pi$}
\label{sec:om3pi}

The differential decay width of the process $\omega \to \pi^+ \, \pi^- \, \pi^0$
is determined by the matrix element
\begin{eqnarray}
  && C_{\omega\to \pi^+ \, \pi^- \, \pi^0} =
  \frac{m_V \, h_P \, h_A}{4 \, f^3 \, m_\omega} \,
  \left[ S_\rho(q_1^2) \, (q_1^2 + m_\omega^2)
  \right. \nonumber \\ && \left. {} \hspace*{5em}
    + S_\rho(q_2^2) \, (q_2^2 + m_\omega^2) + S_\rho(q_3^2) \, (q_3^2 + m_\omega^2)  \right]
  \nonumber \\ && {}
  - \frac{2\, m_V \, h_P \, m_\pi^2 \, b_A}{f^3 \, m_\omega} \,
  \left[ S_\rho(q_1^2) + S_\rho(q_2^2) + S_\rho(q_3^2) \right]   \,,
  \label{eq:C-om3pi}
\end{eqnarray}
where we use the kinematics of (\ref{eq:two-body-var}). In (\ref{eq:C-om3pi}) the rho-meson propagator
\begin{equation}
  \label{eq:def-rhoprop}
  S_\rho(q^2) = \frac{1}{q^2 - m_\rho^2 + i \Gamma_\rho(q^2)}
\end{equation}
appears. Here $\Gamma_\rho(q^2)$ denotes the energy-dependent width of the rho meson. It turns
out that the results only change by about 2\%, if this width is neglected. We do not think that our
leading-order calculation and the determination of our coupling constants have a comparable accuracy.
Hence we refrain from a more detailed modeling of the rho-meson propagator including e.g.\ an energy
dependent real part of the self energy.

The factors $q_i^2+m_\omega^2$ which accompany the rho-meson propagators in (\ref{eq:C-om3pi}) emerge from
the use of the tensor representation for the vector mesons. In other representations this combination
might split up into contact terms and vector-meson exchange terms. In our approach we do not have extra
contact terms in leading order and therefore one parameter combination $\sim h_P \, h_A$ determines the
decay process, at least in the chiral limit.
Quantitatively the extra term $\sim m_\pi^2 \, b_A$ plays only a minor role.

\begin{figure}[ht]
\resizebox{0.48\textwidth}{!}{%
    \includegraphics{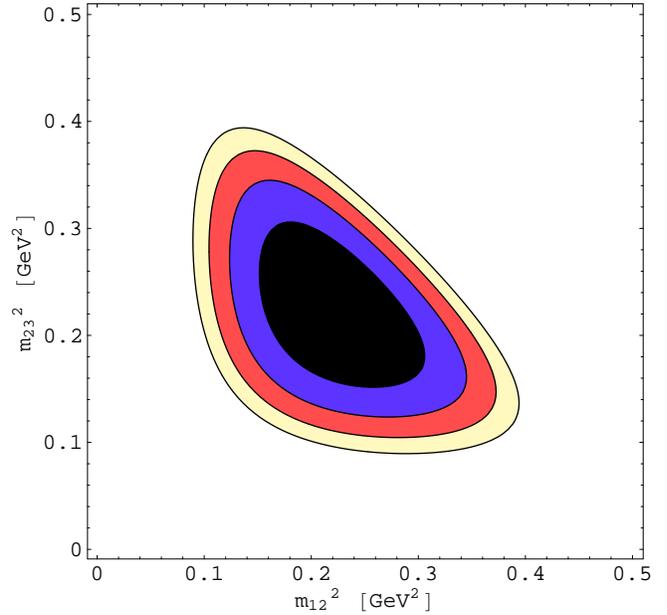}
}
  \caption{Dalitz plot for the decay process $\omega\to \pi^+ \, \pi^- \, \pi^0$
    (arbitrary normalization). Note $m^2_{12}= m^2_{\pi^+\pi^-}$ and $m^2_{23}=m^2_{\pi^- \pi^0}$.}
  \label{fig:dalitz-om}
\end{figure}

\begin{figure}[ht]
\resizebox{0.48\textwidth}{!}{%
    \includegraphics{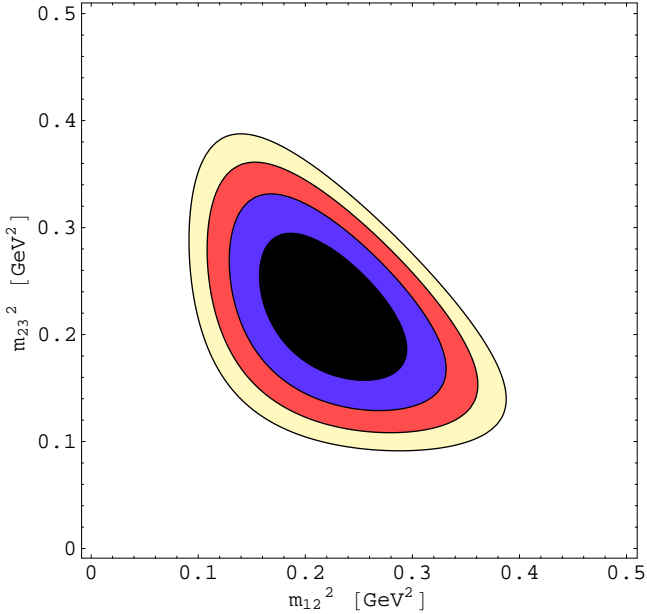}
}
  \caption{Dalitz plot for the phase space of the
    decay process $\omega\to \pi^+ \, \pi^- \, \pi^0$. }
  \label{fig:phasespace-om}
\end{figure}

In fig.\ \ref{fig:dalitz-om} the Dalitz plot for the decay
process $\omega\to \pi^+ \, \pi^- \, \pi^0$ is depicted using the
values (\ref{eq:paramval}). The shape can be compared
with the one obtained from pure phase space, i.e.\ replacing
$C_{\omega\to \pi^+ \, \pi^- \, \pi^0} $ by a constant. A pure phase space Dalitz shape
is shown in fig.\ \ref{fig:phasespace-om}. There are no visible differences and
indeed in the kinematically allowed range the squared matrix element
$\vert C_{\omega\to \pi^+ \, \pi^- \, \pi^0} \vert^2$ is rather flat (not shown).

\begin{figure}[ht]
\resizebox{0.48\textwidth}{!}{%
    \includegraphics{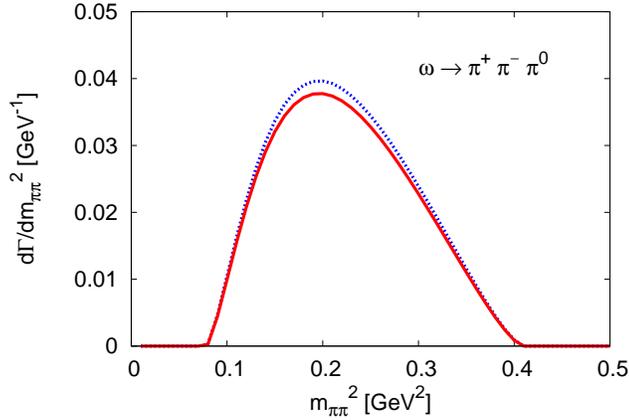}
}
  \caption{Single-differential decay width for the
    process $\omega\to \pi^+ \, \pi^- \, \pi^0$. The solid line gives the full result, the dotted with 
    $b_A=0$. }
  \label{fig:singlediff-om}
\end{figure}

The single-differential decay width $\displaystyle\frac{d\Gamma_{\omega\to 3\pi}}{dm_{12}^2}$ is plotted
in fig.\ \ref{fig:singlediff-om}. One observes the typical rise and fall of phase space.
We predict that the
shape of the fully and single-differential decay widths do not differ significantly from pure
phase space. It was checked that this finding emerges from the interference of the respective three
rho-meson terms which appear in (\ref{eq:C-om3pi}). Physically this corresponds to the three decays
$\omega \to \rho^+ \, \pi^-$, $\omega \to \rho^0 \, \pi^0$, $\omega \to \rho^- \, \pi^+$.
If there was only one rho meson, e.g.\ the $\rho^+$, the Dalitz shape would differ significantly from
pure phase space.

For the partial decay width one gets
\begin{equation}
  \label{eq:omegawidththeo}
  \Gamma_{\omega\to 3\pi} = 7.3 \, \mbox{MeV}
\end{equation}
which agrees very well with the experimental value \cite{Yao:2006px}
\begin{equation}
  \label{eq:omegawidthexp}
  \Gamma_{\omega\to 3\pi}^{\rm exp} = (7.57 \pm 0.13) \, \mbox{MeV.}
\end{equation}
As stated before the influence of $b_A $ is of minor importance here. Using $b_A=0$ leads to 
$\Gamma_{\omega\to 3\pi} = 7.7$ MeV. We emphasize that our coupling constants have been 
obtained from other processes. Thus, the good agreement
between (\ref{eq:omegawidththeo}) and (\ref{eq:omegawidthexp}) can be regarded as a further justification
for our approach which predicts that the considered three-body decay processes are governed by vector-meson
dominance (cf.\ fig.\ \ref{fig:vto3p}).
We stress again that such a statement hinges on the representation used for the vector meson
fields (cf.\ also the discussion in \cite{Ecker:1989yg}). In our work we use tensor fields and our
counting scheme leaves no room for contact terms at leading order. Other works, which use the vector
representation, do find additional contact terms, see e.g.\ \cite{Klingl:1996by} and references therein.
We stress again that in these works the experimental decay width (\ref{eq:omegawidthexp}) is used to
pin down the coupling constants of the respective model. In contrast, our result (\ref{eq:omegawidththeo})
is a consequence of our effective Lagrangian which connects in a systematic way the two-body decays studied
in \cite{Lutz:2008km} with the three-body decays studied here.

\section{Decay $K^* \to K\,\pi\,\pi$}
\label{sec:dec-KsKpipi}

We turn to the decay modes of the ${K^*}^+$. Due to isospin symmetry the decay modes
of the ${K^*}^0$ need not be considered separately. Three decay channels exist:
${K^*}^+ \to \pi^+ \, \pi^- \, K^+$, ${K^*}^+ \to \pi^0 \, K^+ \, \pi^0$,
${K^*}^+ \to \pi^+ \, K^0 \, \pi^0$.

Let us first discuss the experimental situation:
In \cite{Jongejans:1977ty} upper limits are given for the
processes
\begin{eqnarray}
 & \Gamma_{{K^*}^- \to \pi^+ \, \pi^- \, K^-}  < & 40 \, \mbox{keV,}
  \nonumber \\
  &\Gamma_{{K^*}^- \to \pi^- \, \pi^0 \, \bar K^0}  < & 35 \, \mbox{keV,}
  \nonumber \\
  &\Gamma_{{\bar K^{*0}} \to \pi^+ \, \pi^- \, \bar K^0}  < & 35 \, \mbox{keV.}
  \label{eq:decKsexp1}
\end{eqnarray}
Using particle-antiparticle symmetry and flipping the iso\-spin for the latter process
this yields the following constraints for the ${K^*}^+$ decays:
\begin{eqnarray}
  & \Gamma_{{K^*}^+ \to \pi^+ \, \pi^- \, K^+}  < & 40 \, \mbox{keV,}
  \nonumber \\
  & \Gamma_{{K^*}^+ \to \pi^+ \, \pi^0 \, K^0}  < & 35 \, \mbox{keV,}
  \nonumber \\
  & \Gamma_{{K^*}^+ \to \pi^+ \, \pi^- \, K^+}  < & 35 \, \mbox{keV.}
  \label{eq:decKsexp2}
\end{eqnarray}

\subsection{Decay ${K^*}^+ \to \pi^+ \, \pi^- \, K^+$}
\label{sec:dec-KsKpipi1}

We identify the three states with momenta
$p_1$, $p_2$, $p_3$ with $\pi^+$, $\pi^-$, $K^+$, respectively. The relevant matrix element to be used in (\ref{eq:diffgam}) 
is 
\begin{eqnarray}
  && C_{{K^*}^+ \to \pi^+ \, \pi^- \, K^+} =
\nonumber \\ &&
  \frac{m_V \, h_P \, h_A}{8 \, f^3 \, m_{K^*}} \,
  \left[ S_\rho(q_3^2) \, (q_3^2 + m_{K^*}^2) +
    S_{K^*}(q_1^2) \, (q_1^2 + m_{K^*}^2)  \right]
\nonumber \\ && {}
  - \frac{m_V \, h_P \, m_\pi^2 \, b_A}{f^3 \, m_{K^*}} \,
  \left[ \frac{m_K^2}{m_\pi^2} \, S_\rho(q_3^2) + S_{K^*}(q_1^2)  \right]   \,.
  \label{eq:C-Ksp-pich}
\end{eqnarray}
The propagator of the $K^*$ meson is constructed in the same way as the rho-meson
propagator (\ref{eq:def-rhoprop}). Also here it does not influence the results
appreciably, if the energy dependent width of the $K^*$ meson is neglected.

\begin{figure}[ht]
\resizebox{0.48\textwidth}{!}{%
    \includegraphics{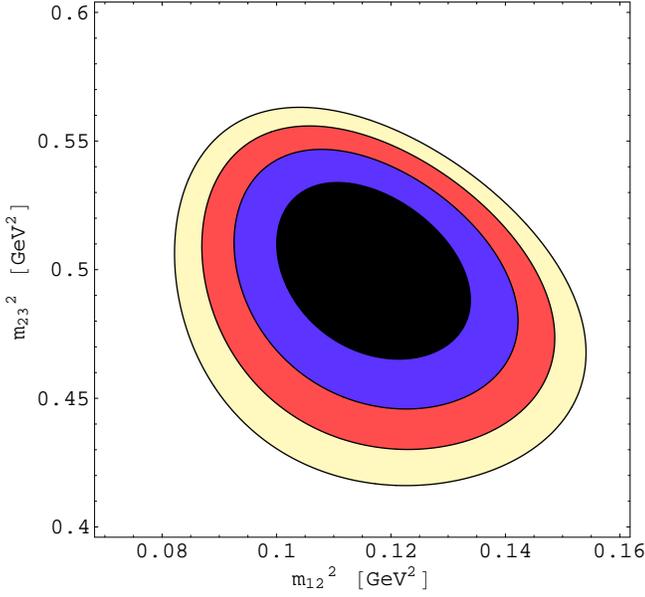}
}
  \caption{Dalitz plot for the decay process ${K^*}^+ \to \pi^+ \, \pi^- \, K^+$
    (arbitrary normalization). Note $m^2_{12}= m^2_{\pi^+\pi^-}$ and $m^2_{23}=m^2_{\pi^- K^+}$.}
  \label{fig:dalitz-Ks1}
\end{figure}

\begin{figure}[ht]
\resizebox{0.48\textwidth}{!}{%
    \includegraphics{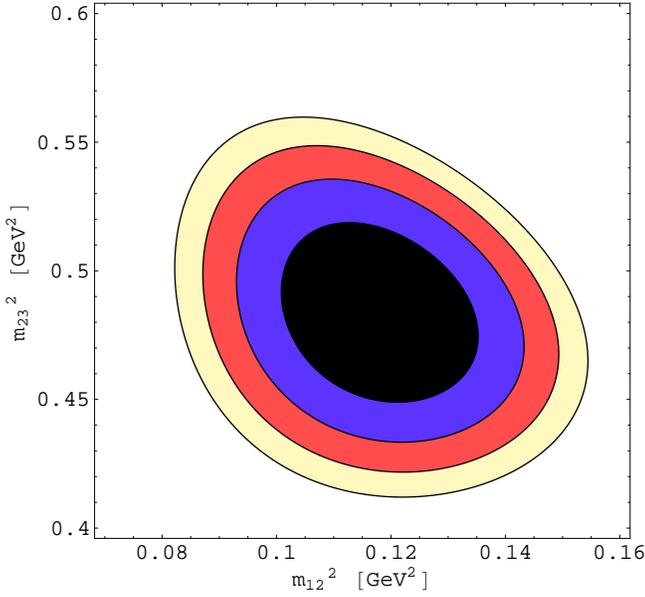}
}
  \caption{Dalitz plot for the phase space of the
    decay process ${K^*}^+ \to \pi^+ \, \pi^- \, K^+$. }
  \label{fig:phasespace-Ks1}
\end{figure}

The Dalitz plot for this reaction is shown in fig.\ \ref{fig:dalitz-Ks1}.
The corresponding plot with pure phase space is depicted in
fig.\ \ref{fig:phasespace-Ks1}. There are some deviations from pure phase space
which are interesting to be resolved experimentally.

\begin{figure}[ht]
\resizebox{0.48\textwidth}{!}{%
    \includegraphics{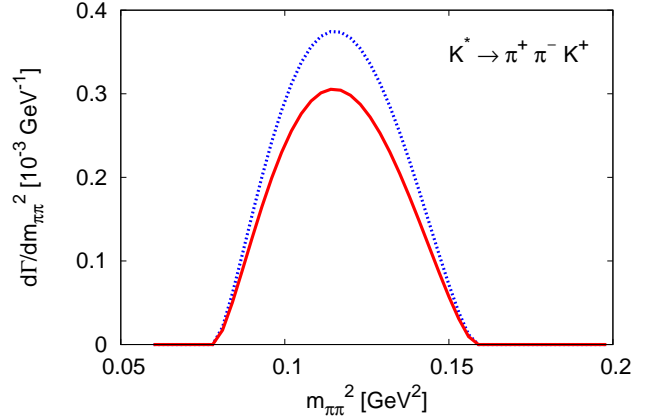}
}
  \caption{Single-differential decay width for the
    process ${K^*}^+ \to \pi^+ \, \pi^- \, K^+$. The solid line gives the full result, the dotted with
    $b_A=0$. }
  \label{fig:singlediff-Ks1}
\end{figure}

The single-differential decay
width $\displaystyle\frac{d\Gamma_{{K^*}^+ \to \pi^+ \, \pi^- \, K^+}}{dm_{12}^2}$ is plotted
in fig.\ \ref{fig:singlediff-Ks1}. No significant deviations from pure phase space remain in this
integrated quantity (not shown).

For the partial decay width one gets
\begin{equation}
  \label{eq:Ks1widththeo}
  \Gamma_{{K^*}^+ \to \pi^+ \, \pi^- \, K^+} = 14 \, \mbox{keV,}
\end{equation}
in agreement with the experimental constraints (\ref{eq:decKsexp2}).  Switching off the effect of $b_A$ increases the width by 
$4$ keV. The order of magnitude of our result (\ref{eq:Ks1widththeo}) agrees with the range given in \cite{Boal:1976eg}.
It clearly would be interesting to confront our prediction with experimental results. In high-statistics experiments like e.g.\
HADES at GSI  and WASA at COSY it might be possible to observe such rare decays.

\subsection{Decay ${K^*}^+ \to \pi^0 \, K^+ \, \pi^0$}
\label{sec:dec-KsKpipi2}

For the decay ${K^*}^+ \to \pi^0 \, K^+ \, \pi^0$ we identify the three states
1, 2, 3 with $\pi^0$, $K^+$, $\pi^0$, respectively. It holds
\begin{eqnarray}
  && C_{{K^*}^+ \to \pi^0 \, K^+ \, \pi^0} =   \frac{m_V \, h_P \, h_A}{16 \, f^3 \, m_{K^*}} \,
\nonumber \\ && \phantom{m}  \times
  \left[ S_{K^*}(q_3^2) \, (q_3^2 + m_{K^*}^2) -
    S_{K^*}(q_1^2) \, (q_1^2 + m_{K^*}^2)  \right]
\nonumber \\ && {}
  - \frac{m_V \, h_P \, m_\pi^2 \, b_A}{2 \, f^3 \, m_{K^*}} \,
  \left[ S_{K^*}(q_3^2) - S_{K^*}(q_1^2)  \right]   \,.
  \label{eq:C-Ksp-pineu}
\end{eqnarray}
In addition, for the decay width one has to account for the presence of two
identical states by a factor $1/2$.

\begin{figure}[ht]
\resizebox{0.48\textwidth}{!}{%
    \includegraphics{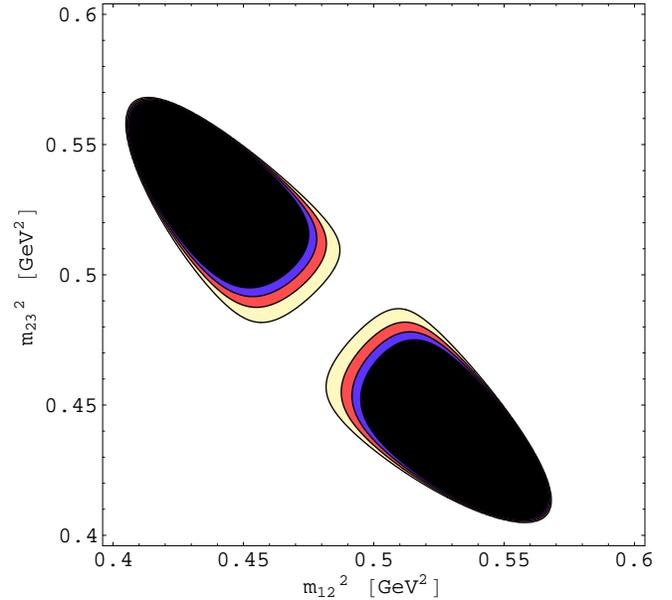}
}
  \caption{Dalitz plot for the decay process ${K^*}^+ \to \pi^0 \, K^+ \, \pi^0$
    (arbitrary normalization). Note $m^2_{12}= m^2_{\pi^0 K^+}$ and $m^2_{23}=m^2_{K^+ \pi^0}$.}
      \label{fig:dalitz-Ks2}
\end{figure}

\begin{figure}[ht]
\resizebox{0.48\textwidth}{!}{%
    \includegraphics{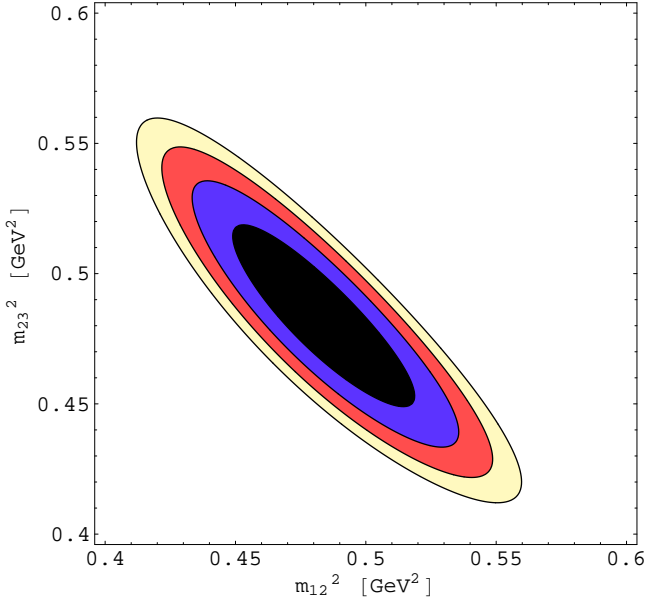}
}
  \caption{Dalitz plot for the phase space of the
    decay process ${K^*}^+ \to \pi^0 \, K^+ \, \pi^0$.}
  \label{fig:phasespace-Ks2}
\end{figure}

The Dalitz plot for this reaction is shown in fig.\ \ref{fig:dalitz-Ks2}.
Due to the destructive interference present in
(\ref{eq:C-Ksp-pineu}) there is a striking difference to pure phase space shown
in fig.\ \ref{fig:phasespace-Ks2}.

\begin{figure}[ht]
\resizebox{0.48\textwidth}{!}{%
    \includegraphics{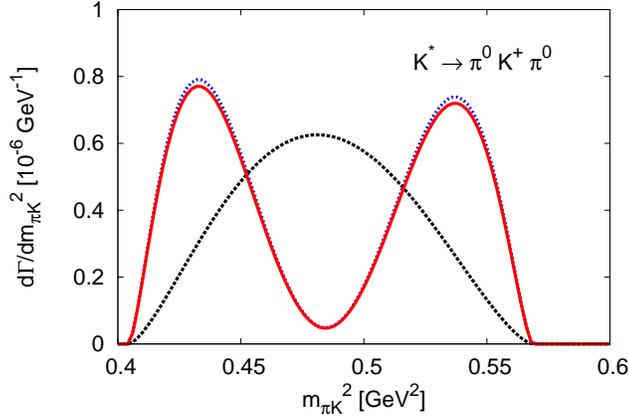}
}
  \caption{Single-differential decay width for the
    process ${K^*}^+ \to \pi^0 \, K^+ \, \pi^0$.   
    The solid line gives the full result, the dotted with
    $b_A=0$. For comparison also the shape for pure
    phase space is plotted as the dashed line.}
  \label{fig:singlediff-Ks2}
\end{figure}

The single-differential decay
width $\displaystyle \frac{d\Gamma_{{K^*}^+ \to \pi^0 \, K^+ \, \pi^0}}{dm_{12}^2}$ is plotted
in fig.\ \ref{fig:singlediff-Ks2}. Again the difference to pure phase space is obvious.
For the partial decay width one gets the value
\begin{equation}
  \label{eq:Ks2widththeo}
  \Gamma_{{K^*}^+ \to \pi^0 \, K^+ \, \pi^0} = 67 \, \mbox{eV} \,,
\end{equation}
whereas $b_A=0$ leads to $69$ eV.  
It has already been pointed out in \cite{Boal:1976eg}
that this particular decay channel has a rather small width.
Of course, this smallness is caused by the already mentioned negative interference. Hence, while
a fully or single-differential width shows an interesting pattern, the overall missing strength of this
decay channel probably inhibits an experimental verification.

\subsection{Decay ${K^*}^+ \to \pi^+ \, K^0 \, \pi^0$}
\label{sec:dec-KsKpipi3}

\begin{figure}[ht]
\resizebox{0.48\textwidth}{!}{%
    \includegraphics{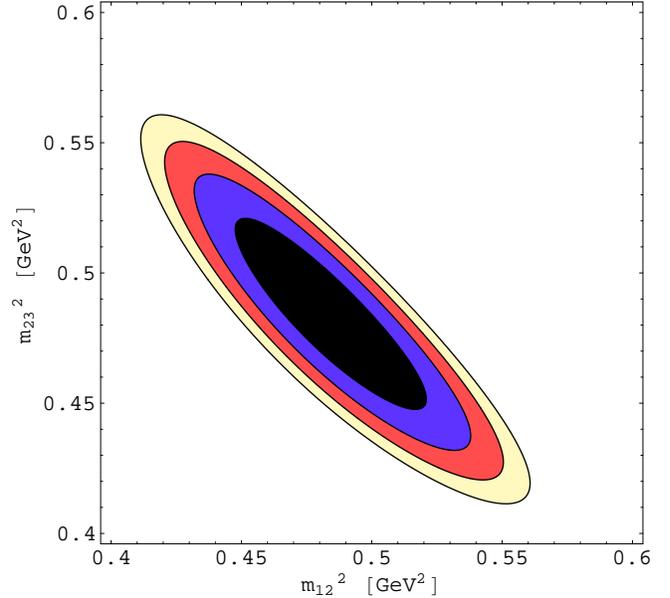}
}
  \caption{Dalitz plot for the decay process ${K^*}^+ \to \pi^+ \, K^0 \, \pi^0$
    (arbitrary normalization). Note $m^2_{12}= m^2_{\pi^+ K^0}$ and $m^2_{23}=m^2_{K^0 \pi^0}$.}
  \label{fig:dalitz-Ks3}
\end{figure}

We identify the three states 1, 2, 3 in (\ref{eq:diffgam}) with $\pi^+$, $K^0$, $\pi^0$, respectively and derive
\begin{eqnarray}
  && C_{{K^*}^+ \to \pi^+ \, K^0 \, \pi^0} =
  \frac{m_V \, h_P \, h_A}{8 \, \sqrt{2} \, f^3 \, m_{K^*}} \,
  \left[ S_{K^*}(q_1^2) \, (q_1^2 + m_{K^*}^2)
\right. \nonumber \\ && \left. \hspace*{3.4em}   {}
    + S_{K^*}(q_3^2) \, (q_3^2 + m_{K^*}^2)
    + 2 \, S_\rho(q_2^2) \, (q_2^2 + m_{K^*}^2)  \right]
\nonumber \\ && {}
  - \frac{m_V \, h_P \, m_\pi^2 \, b_A}{\sqrt{2} \, f^3 \, m_{K^*}} \,
  \left[ S_{K^*}(q_1^2) + S_{K^*}(q_3^2) + 2 \, \frac{m_K^2}{m_\pi^2} \, S_\rho(q_2^2) \right]   \,.
\nonumber \\ &&
  \label{eq:C-Ksp-K0}
\end{eqnarray}
The Dalitz plot for this reaction is shown in fig.\ \ref{fig:dalitz-Ks3}.
In practice there is no difference to pure phase space shown in
fig.\ \ref{fig:phasespace-Ks2}.

\begin{figure}[ht]
\resizebox{0.48\textwidth}{!}{%
    \includegraphics{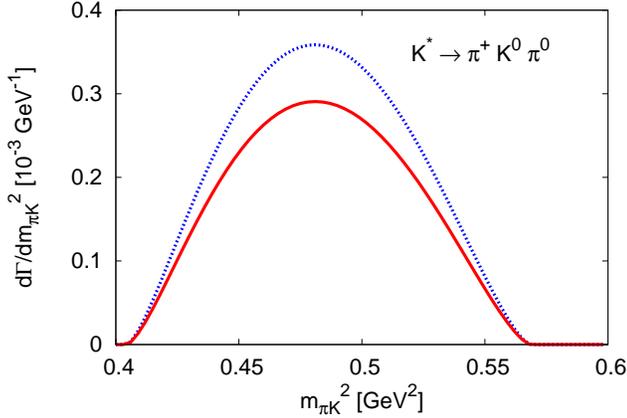}
}
  \caption{Single-differential decay width for the
    process ${K^*}^+ \to \pi^+ \, K^0 \, \pi^0$. The solid line gives the full result, the dotted with
    $b_A=0$. }
  \label{fig:singlediff-Ks3}
\end{figure}

The single-differential decay
width $\displaystyle \frac{d\Gamma_{{K^*}^+ \to \pi^+ \, K^0 \, \pi^0}}{dm_{12}^2}$ is plotted
in fig.\ \ref{fig:singlediff-Ks3}.
For the partial decay width one gets
\begin{equation}
  \label{eq:Ks3widththeo}
  \Gamma_{{K^*}^+ \to \pi^+ \, K^0 \, \pi^0} = 28 \, \mbox{keV.}
\end{equation}
This result is in agreement with the experimental constraints (\ref{eq:decKsexp2}).
We observe that for this decay the influence of $b_A$ is significant. For $b_A=0$ the decay width 
comes at $35$ keV, right at the boundary of  the empirical constraint (\ref{eq:decKsexp2}). 
Qualitatively, also in \cite{Boal:1976eg} it was found that the
decay width (\ref{eq:Ks3widththeo}) is about twice as large as the one given
in (\ref{eq:Ks1widththeo}).

Among the three-body decay branches of the ${K^*}^+$ meson the partly
charged channel $\pi^+ \, K^0 \, \pi^0$ has the largest width. An experimental verification of
our results (\ref{eq:Ks1widththeo}), (\ref{eq:Ks2widththeo}), (\ref{eq:Ks3widththeo})
for the partial widths and maybe even for the differential widths shown in
figs.\ \ref{fig:dalitz-Ks1}, \ref{fig:singlediff-Ks1}, \ref{fig:dalitz-Ks2}, \ref{fig:singlediff-Ks2},
\ref{fig:dalitz-Ks3} and \ref{fig:singlediff-Ks3} would be highly desirable. Like in the radiative two-body decays
studied in \cite{Lutz:2008km} we predict sizeable flavor breaking effects in the hadronic three-body decays. 

\section{Summary}
\label{sec:sum}

The chiral Lagrangian with Goldstone bosons and light vector mesons  
has been used to calculate hadronic three-body decays of light vector mesons following the recently proposed counting 
rules \cite{Lutz:2008km}. The leading order results agree very well with all available data and constraints. We have 
presented predictions for differential decay widths of all considered processes and for the partial decay widths of the
three-body decays of the $K^*$ meson. We suggest to study such reactions in high-statistics experiments like 
HADES at GSI  and WASA at COSY.


\bibliographystyle{epj}
\bibliography{literature-m}

\end{document}